# Ferromagnetic phase of a uniaxial magnet with anisotropic biquadratic exchange

**I.P. SHAPOVALOV**

I.I. Mechnikov Odesa National University
*(2, Dvoryans'ka Str., Odesa 270100, Ukraine)*

The ferromagnetic phase (FMP) of a uniaxial magnet with the easy-plane single-ion anisotropy (SIA) and the anisotropic biquadratic exchange interaction (BQEI) has been studied. The case $S = 1$ for the site spin $S$ has been considered. Expressions for two branches of the spin excitation spectrum at finite temperatures $T$ have been obtained, and the conditions for spectral mode stability have been determined. The spectral mode stability diagram in the $T-h$ coordinates has been constructed. The diagram testifies that, under certain conditions, the temperature decrease is accompanied by a violation of the spectral mode stability followed, as the temperature decreases further, by its restoration; i.e. the reentrance phenomenon is observed. The temperature of the second-order phase transition (PT) from the FMP into the phase with spontaneously broken symmetry has been demonstrated to depend considerably on the BQEI anisotropy constant.

## 1. Introduction

Magnets with high values of the SIA and BQEI constants have been found in works [1–6]. In turn, it invoked the subsequent researches of such systems [7–23]. However, in the majority of works, where the BQEI was considered, the authors confined themselves to the isotropic BQEI approximation.

In work [17], magnets with SIA and anisotropic BQEI with the site spin $S = 1$ were studied. It was found that, in the case where the external magnetic field is directed along the crystal symmetry axis (the $z$-axis), two phases with spontaneously broken symmetry can be realized in the system, besides the symmetric ferromagnetic and quadrupole phases. One of the asymmetric phases is the so-called $Q_<FM_Z$ phase. It is a quadrupole-ferromagnetic phase with both a ferromagnetic ordering axis coinciding with the $z$-axis and a plane of quadrupole ordering, whose orientation depends on Hamiltonian parameters. When the magnetic field $h_Z$ grows, the fraction of the ferromagnetic component increases and that of the quadrupole one decreases. At a definite $h_Z$-value, the phase $Q_<FM_Z$ continuously transforms into the FMP, i.e. there occurs a PT of the second kind induced by the field. Unfortunately, the authors of work [17] confined the consideration to the low-temperature case, which made the study of temperature-induced PTs impossible. A generalization of the ferromagnetic and $Q_<FM_Z$ phase researches to the finite temperature interval has been carried out in work [20]. In particular, a boundary between the ferromagnetic and $Q_<FM_Z$ phases in the field versus temperature coordinates was determined. However, the problem of spin excitation spectra remained unresolved.

Another asymmetric phase (the $Q_<FM_<$ phase, according to the terminology of the authors of work [17]), is a phase, in which the magnetization is directed at an angle to the field $h_Z$. At $T = 0$, when the external field grows, the quadrupole phase (QP) → $Q_<FM_<$ phase transition occurs at $h_Z = h_{c1}$. If the field grows further, the phase transition, either $Q_<FM_<$ phase → $Q_<FM_Z$ phase or $Q_<FM_<$ phase → FMP, occurs at $h_Z = h_{c2}$. Hence, the $Q_<FM_<$ phase is realized only provided that $h_{c1} < h_{c2}$, with $h_{c1}$- and $h_{c2}$-values depending on Hamiltonian parameters. In work [24], expressions for $h_{c1}$ and $h_{c2}$ were obtained in the case where there exists anisotropic BQEI in the system (the corresponding expressions are given below).

In the absence of BQEI, only one asymmetric phase can be realized in an easy-plane magnet, namely, in the case where the magnetic field is perpendicular to the easy plane. This phase is an analog of the $Q_<FM_<$ one (an angular phase). The existence of this phase was predicted in works [25, 26] and confirmed by experiments carried out with nickel compounds [27–30]. Further researches of the angular phase have been continued till now [32–35]. For instance, in work [33], an experimental $T - h$ phase diagram is presented, in which the boundaries of the angular phase with the quadrupole and ferromagnetic ones agree well with the experimental results of work [31]. A comparison between the results obtained in works [33] and [24] gives the following result: the expressions for $h_{c1}$ obtained in both works completely agree with each other, whereas expressions for $h_{c2}$ are different to a certain extent.

The spectra of spin excitations in magnets with anisotropic BQEI at finite temperatures were the object of researches in work [36], where a special method was developed. The method is based on the application of a suitable dynamic matrix, the characteristic values of which coincide with the values for spin excitation energies.

The main purpose of this work consists in constructing the spin excitation spectra for a FMP and determining the conditions of their stability. The analysis of the problem is carried out within the method elaborated in work [36].

## 2. Hamiltonian

In the most general case, the uniaxial Hamiltonian with $S = 1$ which makes allowance for SIA and BQEI looks like

$$H = -h_Z \sum_i S_i^Z -$$

$$- \sum_{i,j(i\neq j)} J_{ij} \left[ S_i^Z S_j^Z - 2\xi S_i^+ S_j^- \right] + D \sum_i O_{2i}^0 -$$

$$- \sum_{i,j(i\neq j)} K_{ij} \left( 3 O_{2i}^0 O_{2j}^O - 2\eta O_{2i}^1 O_{2j}^{-1} + 4\zeta O_{2i}^2 O_{2j}^{-2} \right), \quad (1)$$

where $J_{ij}$ are the exchange interaction constants, $K_{ij}$ are the BQEI constants, $D$ is the SIA constant; $\xi$, $\eta$, and $\zeta$ are positive numbers; and $O_l^m$ with $l = 1, 2$ and $m = 0, \pm 1, \ldots, \pm l$ are the tensor operators that form the Lie algebra of the group SU(3). The first term in Hamiltonian (1) is the energy of site spins in an external magnetic field (the Zeeman energy). The second term is the energy of exchange interaction which becomes isotropic at $\xi = 1$. The third term is the energy of spins in the crystalline field. The fourth term is the BQEI energy. In the case where $\eta = \zeta = 1$, the BQEI is isotropic:

$$H_{\text{BQEI}} = - \sum_{i,j(i\neq j)} K_{ij} \left( S_i S_j \right)^2. \quad (2)$$

The deviations of the parameters $\eta$ and $\zeta$ from 1 characterize the anisotropy degree of BQEI; i.e. those parameters are the BQEI anisotropy constants.

The operators $O_l^m$ are connected with spin operators by the relations

$$O_1^0 = S^Z; \quad O_1^1 \equiv S^+ = \frac{1}{\sqrt{2}} \left( S^X - iS^Y \right);$$

$$O_1^{-1} \equiv S^- = \frac{-1}{\sqrt{2}} \left( S^X + iS^Y \right);$$

$$O_2^0 = \left( S^Z \right)^2 - \frac{2}{3}; \quad O_2^{\pm 1} = - \left( S^Z S^\pm + S^\pm S^Z \right);$$

$$O_2^{\pm 2} = \left( S^\pm \right)^2. \quad (3)$$

The average values of these operators determine the spin ordering in the system. In the FMP, only the diagonal averages $\langle S^Z \rangle$ and $\langle O_2^0 \rangle$ are different from zero, and the order parameter is therefore two-component. In this work, we confine ourselves to the consideration of the single-sublattice ordering in easy-plane magnets, which is provided by the conditions $J_{ij} > 0$, $K_{ij} > 0$, and $D > 0$.

In the molecular-field approximation,

$$H_0 = - \left( h_Z + 2J_0 \langle S^Z \rangle \right) \times$$

$$\times \sum_i S_i^Z + \left( D - 6K_0 \langle O_2^0 \rangle \right) \sum_i O_{2i}^0, \quad (4)$$

where $J_0 \equiv \sum_i J_{ij}$ and $K_0 \equiv \sum_i K_{ij}$.

Depending on the spin projection $S^Z$ on the $z$-axis ($S^Z = 0, \pm 1$), the energy levels of lattice site atoms are determined by the formulas

$$E_0 = -\frac{2}{3} D + 4 K_0 \langle O_2^0 \rangle;$$

$$E_1 = -h_Z - 2 J_0 \langle S^Z \rangle + \frac{1}{3} D - 2 K_0 \langle O_2^0 \rangle;$$

$$E_{-1} = h_Z + 2 J_0 \langle S^Z \rangle + \frac{1}{3} D - 2 K_0 \langle O_2^0 \rangle. \quad (5)$$

Since the condition $E_1 < E_{-1}$ is satisfied automatically in the FMP, this phase can be realized, provided that $E_1 < E_0$ or

$$h_Z + 2 J_0 \langle S^Z \rangle > D - 6 K_0 \langle O_2^0 \rangle. \quad (6)$$

At zero temperature ($T = 0$) in the FMP,

$$\langle S^Z \rangle = 1; \quad \langle O_2^0 \rangle = \frac{1}{3}. \quad (7)$$

At finite temperatures, the averages $\langle S^Z \rangle$ and $\langle O_2^0 \rangle$ are determined by the system of two equations [36]

$$\langle S^Z \rangle = \frac{2\text{sh}\dfrac{h_z + 2J_0 \langle S^Z \rangle}{\theta} \exp\dfrac{6K_0 \langle O_2^0 \rangle - D}{\theta}}{1 + 2\text{ch}\dfrac{h_z + 2J_0 \langle S^Z \rangle}{\theta} \exp\dfrac{6K_0 \langle O_2^0 \rangle - D}{\theta}},$$

$$\langle O_2^0 \rangle = \frac{1}{3} - \frac{1}{1 + 2\text{ch}\dfrac{h_z + 2J_0\langle S^Z \rangle}{\theta} \exp\dfrac{6K_0\langle O_2^0 \rangle - D}{\theta}},\tag{8}$$

where $\theta$ is the temperature expressed in energy units ($\theta = kT$).

Note that system (8) looks identically for the quadrupole and ferromagnetic phases; however, the quantities $\langle S_Z \rangle$ and $\langle O_2^0 \rangle$ are different in both phases. For the identification of solutions of the system which correspond to different phases, it is expedient to use the passage to the limit $T \to 0$. In this case, $\langle S^Z \rangle \to 0$ and $\langle O_2^0 \rangle \to -\frac{2}{3}$ in the QP, and $\langle S^Z \rangle \to 1$ and $\langle O_2^0 \rangle \to \frac{1}{3}$ in the FMP.

## 3. Spin Excitation Spectrum

The value $S^Z = 1$ corresponds to the ground state of site atoms in the FMP. At finite temperatures, there arise the spin excitations with $S^Z = 0$ and $S_Z = -1$. The creation operators for these excitations are the Hubbard operators $X^{01}$ and $X^{-11}$, and the corresponding annihilation operators are $X^{10}$ and $X^{1-1}$.

In order to find branches of the spin excitation spectrum, we used the method proposed in work [36] which consists, briefly, in the following. Calculating the commutators of non-diagonal Hubbard operators and the Hamiltonian, the dynamic matrix is constructed. The number of characteristic values of this matrix coincides with the number of non-diagonal Hubbard operators that were used, with every characteristic value corresponding to a certain Hubbard operator. The expressions for those characteristic values, which correspond to the annihilation operators $X^{01}$ and $X^{-11}$, coincide in the $k$-space with the expressions for branches of the spin excitation spectrum.

To calculate the commutators, it is expedient to pass to Hubbard operators in Hamiltonian (1). At $S^Z = 1$, the relations between the $O_l^m$ and Hubbard operators are given by the formulas

$$S^Z = X^{11} - X^{-1-1}, \quad S^+ = -X^{10} - X^{0-1},$$

$$S^- = X^{01} + X^{-10}, \quad O_2^0 = X^{11} + X^{-1-1} - \frac{2}{3},$$

$$O_2^1 = X^{10} - X^{0-1}, \quad O_2^{-1} = -X^{01} + X^{-10},$$

$$O_2^2 = X^{1-1}, \quad O_2^{-2} = X^{-11}.\tag{9}$$

Accordingly, Hamiltonian (1) reads

$$H = -h_Z \sum_i \left( X_i^{11} - X_i^{-1-1} \right) -$$

$$- \sum_{i,j} J_{ij} \left[ \left( X_i^{11} - X_i^{-1-1} \right)\left( X_j^{11} - X_j^{-1-1} \right) + \right.$$

$$+ 2\xi \left( X_i^{10} + X_i^{0-1} \right)\left( X_j^{01} + X_j^{-10} \right) \Big] +$$

$$+ D \sum_i \left( X_i^{11} + X_i^{-1-1} - \frac{2}{3} \right) -$$

$$- \sum_{i,j} K_{ij} \left[ 3 \left( X_i^{11} + X_i^{-1-1} - \frac{2}{3} \right) \times \right.$$

$$\times \left( X_j^{11} + X_j^{-1-1} - \frac{2}{3} \right) -$$

$$- 2\eta \left( X_i^{10} - X_i^{0-1} \right)\left( -X_j^{01} + X_j^{-10} \right) + 4\zeta X_i^{1-1} X_j^{-11} \Big].\tag{10}$$

To construct a suitable dynamic matrix, it is necessary to calculate the commutators $[X_f^{10}, H]$, $[X_f^{0-1}, H]$, and $[X_f^{1-1}, H]$ and use the approximation

$$X_i X_j = X_i \langle X_j \rangle + \langle X_i \rangle X_j,$$

$$\langle X^{nm} \rangle = 0, \ (n \ne m) \tag{11}$$

which was proposed in work [36]. In the $k$-space, the corresponding commutators are

$$\left[ X_k^{10}, H \right] = p_{11}(\mathbf{k}) X_k^{10} + p_{12}(\mathbf{k}) X_k^{0-1},$$

$$\left[ X_k^{0-1}, H \right] = p_{21}(\mathbf{k}) X_k^{10} + p_{22}(\mathbf{k}) X_k^{0-1},$$

$$\left[ X_k^{1-1}, H \right] = p_{33}(\mathbf{k}) X_k^{1-1},\tag{12}$$

where the coefficients $p_{im}$ constitute a dynamic matrix with the components

$$p_{11}(\mathbf{k}) = h_Z + 2J_0\langle S^Z \rangle - D + 6K_0\langle O_2^0 \rangle -$$

$$- \left( \langle S^Z \rangle + 3\langle O_2^0 \rangle \right)\left( \xi J_k + \eta K_k \right),$$

$$p_{12}(\mathbf{k}) = (\langle S^Z \rangle + 3\langle O_2^0 \rangle)(\eta K_k - \xi J_k),$$

$$p_{21}(\mathbf{k}) = (\langle S^Z \rangle - 3\langle O_2^0 \rangle)(\eta K_k - \xi J_k),$$

$$p_{22}(\mathbf{k}) = h_Z + 2J_0 \langle S^Z \rangle + D - 6K_0 \langle O_2^0 \rangle -$$

$$- (\langle S^Z \rangle - 3\langle O_2^0 \rangle)(\xi J_k + \eta K_k),$$

$$p_{33}(\mathbf{k}) = 2h_Z + 4\langle S^Z \rangle (J_0 - \zeta K_k). \tag{13}$$

The characteristic values of the dynamic matrix are determined by the secular equation

$$\begin{vmatrix} p_{11}(\mathbf{k}) - \varepsilon(\mathbf{k}) & p_{12}(\mathbf{k}) & 0 \\ p_{21}(\mathbf{k}) & p_{22}(\mathbf{k}) - \varepsilon(\mathbf{k}) & 0 \\ 0 & 0 & p_{33}(\mathbf{k}) - \varepsilon(\mathbf{k}) \end{vmatrix} = 0. \tag{14}$$

Its solution gives three characteristic values which can be expressed, by taking Eq. (13) into account, as follows:

$$\varepsilon_1(\mathbf{k}) = 2h_Z + 4\langle S^Z \rangle (J_0 - \zeta K_k), \tag{15}$$

$$\varepsilon_{2,3}(\mathbf{k}) = h_Z + \langle S^Z \rangle (2J_0 - \xi J_k - \eta K_k) \mp$$

$$\mp \left\{ (\langle S^Z \rangle)^2 (\xi J_k - \eta K_k)^2 + \right.$$

$$\left. + [D - 6\langle O_2^0 \rangle (K_0 - \xi J_k)][D - 6\langle O_2^0 \rangle (K_0 - \eta K_k)] \right\}^{1/2}. \tag{16}$$

Those characteristic values of the dynamic matrix, which correspond to the Hubbard annihilation operators, coincide with the branches of the spin excitation spectrum. The characteristic values $\varepsilon_1(k)$ and $\varepsilon_2(k)$ correspond to the operators $X^{10}$ and $X^{1-1}$; therefore, the branches of the spin excitation spectrum are

$$\omega_1(\mathbf{k}) = \varepsilon_1(\mathbf{k}), \tag{17}$$

$$\omega_2(\mathbf{k}) = \varepsilon_2(\mathbf{k}). \tag{18}$$

Since $J_k$ and $K_k$ are even functions of the wave vector $\mathbf{k}$, both branches are characterized by the square-law dispersion in the long-wave limit,

$$\omega_1(\mathbf{k}) = \Delta_1 + \alpha_1 \mathbf{k}^2; \quad \omega_2(\mathbf{k}) = \Delta_2 + \alpha_2 \mathbf{k}^2. \tag{19}$$

In work [37], the inequality $\omega(k) > \omega(0)$ was proved to be valid at every $k \neq 0$ in the cases where the single-sublattice ordering takes place in the system. Therefore, the condition for spectrum mode stability is given by the system of inequalities

$$\omega_1(0) > 0,$$
$$\omega_2(0) > 0, \tag{20}$$

and the stability boundary is determined by two equalities, $\omega_1(0) = 0$ and $\omega_2(0) = 0$, or

$$h_Z + 2\langle S^Z \rangle (J_0 - \zeta K_0) = 0, \tag{21}$$

$$h_Z + \langle S^Z \rangle (2J_0 - \xi J_0 - \eta K_0) =$$

$$= \left\{ (\langle S^Z \rangle)^2 (\xi J_k - \eta K_k)^2 + [D - 6\langle O_2^0 \rangle (K_0 - \xi J_k)] \times \right.$$

$$\left. \times [D - 6\langle O_2^0 \rangle (K_0 - \eta K_k)] \right\}^{1/2}. \tag{22}$$

Expression (22) coincides with the corresponding expression for the QP (see work [36]). However, since the quantities $\langle S^Z \rangle$ and $\langle O_2^0 \rangle$ are different in the ferromagnetic and quadrupole phases, the stability boundaries in the $T - h$ coordinates do not coincide, generally speaking, in both phases.

Note that curve (21) coincides with the curve corresponding to the second-kind PT between the Q$_<$FM$_Z$ and ferromagnetic phases, which was obtained in work [20] (see Introduction).

## 4. Stability Diagram

First of all, it is worth noting that we consider the case where the phase Q$_<$FM$_<$ is not realized, i.e. the condition $h_{c1} > h_{c2}$ is satisfied. For the critical fields $h_{c1}$ and $h_{c2}$, the following expressions were obtained in work [24]:

$$h_{c1} = \sqrt{[D + 4K_0(1-\eta)][D + 4(K_0 - \xi J_0)]},$$

$$h_{c2} = D - 2J_0(1-\xi) - 2K_0(1-\eta). \tag{23}$$

In Fig. 1, the mode stability diagram for the FMP spectrum is depicted in the $\tilde{\theta} - \tilde{h}$ coordinates, where $\tilde{\theta} = \theta/K_0$ and $\tilde{h} = h_Z/K_0$. The Hamiltonian parameters are chosen so that the condition $h_{c1} > h_{c2}$ is satisfied, i.e. a unique asymmetric phase is the phase Q$_<$FM$_<$. Zone *1* is the range, where FMP spectrum modes are stable,

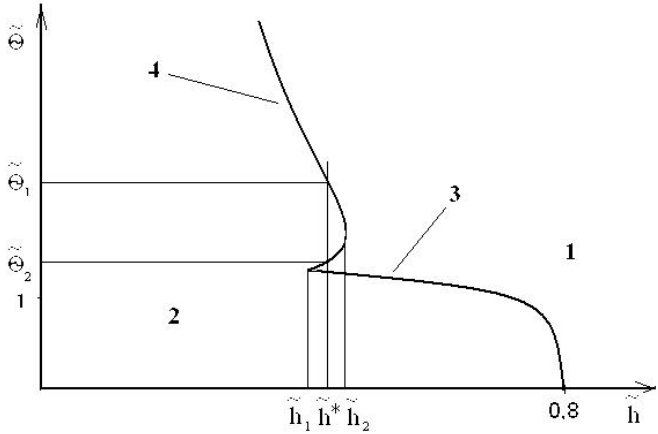

Fig. 1. Mode stability diagram for the FMP spectrum of an easy-plane magnet with anisotropic BQEI: (*1*) region of FMP spectrum mode stability, (*2*) region of FMP spectrum mode instability, (*3*) curve given by expression (21), and (*4*) curve given by expression (22). The diagram is plotted for the parameter values $J_0 = 0.8$, $D = 0.4$, $K_0 = 1$, $\zeta = 1.2$, $\eta = 0.8$, and $\xi = 1.25$

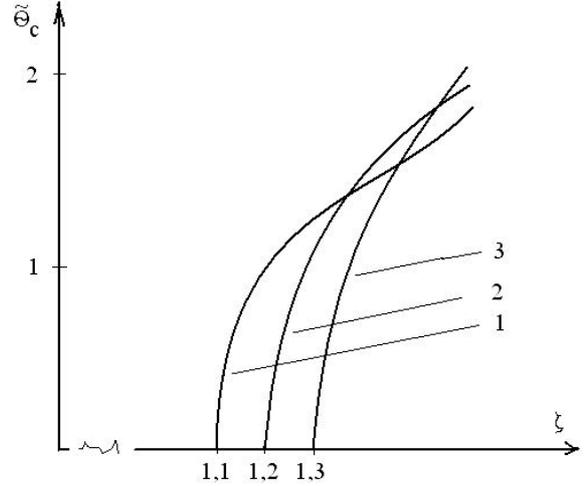

Fig. 2. Dependences of the temperature of the second-kind PT between the ferromagnetic and $Q_<FM_Z$ phases on the constant $\zeta$ at various $h_Z = 0.6$ (*1*), 0.8 (*2*), and 1 (*3*). All curves were calculated for the parameter values $J_0 = 0.8$, $D = 0.4$, and $K_0 = 1$

and zone *2* is the range, where their stability is violated. Curves *3* and *4* are given by expressions (21) and (22), respectively. In the external magnetic field interval $\tilde{h}_1 < \tilde{h} < \tilde{h}_2$, the stability of spectrum modes demonstrates a reentrant behavior. In particular, at $\tilde{h} = \tilde{h}^*$, if the temperature decreases, the stability is first violated at the point $\tilde{\theta}_1$. But the further temperature decrease gives rise to the restoration of the spectrum mode stability at the point $\tilde{\theta}_2$.

Since curve (21) is not only a curve, where the mode stability of the spin excitation spectrum is violated, but also a PT curve, it is expedient to study the dependence of the temperature $\tilde{\theta}_c$ of the second-kind PT between the $Q_<FM_Z$ and ferromagnetic phases on the BQEI anisotropy constant $\zeta$ at various fields $\tilde{h}$. Such a dependence is depicted in Fig. 2. The figure demonstrates that the transition temperature substantially depends on the constant $\zeta$. At the same time, for large enough $\zeta$'s, the temperature $\tilde{\theta}_c$ is almost independent of the external field $\tilde{h}$.

## 5. Discussion of Results

In this work, the expressions for two branches of the spin excitation spectrum in the FMP have been obtained. Both branches demonstrate the square law of the dispersion in the long-wave limit. When determining the spin excitation spectrum branches, the condition $D > 0$ was not used. Therefore, expressions (17) and (18) for the spectrum branches and expressions (21) and (22) for the boundary of the spectrum mode stability range remain valid in the case $D < 0$, i.e. for an easy-axis magnet.

Owing to a mismatch between the stability boundaries for the ferromagnetic and quadrupole phases (see Section 3), there emerges a region in the stability diagram plotted in the $T - h$ coordinates, where the modes of spectra of both phases are stable. The presence of such a region brings about two essential consequences. First, the curve of the PT between the ferromagnetic and quadrupole phases coincides with the curve, where the free energies in both phases are identical; in this case, the corresponding PT is of the first kind. Second, there are two regions of metastability. In one of them, the FMP is metastable and the QP is stable; in the other, the situation is opposite. Hence, the stability diagram does not coincide with the phase one. The results of researches of phase diagrams, metastable regions, and the influence of the BQEI anisotropy constants on the first-kind PT between the ferromagnetic and quadrupole phases will be reported elsewhere.

It has to be noted that the method proposed in work [36] can be directly used only for those phases which preserve the Hamiltonian symmetry, i.e. for the QP and the FMP. In the case of phases with spontaneously broken symmetry, it is necessary, first, to diagonalize the zero Hamiltonian with the help of a unitary transformation for the application of the method proposed to be eligible. This transformation is one-parametric for the $Q_<FM_Z$ phase and two-parametric for the $Q_<FM_<$ one. The calculations of spin excitation spectra in the asym-metric phases at $T \neq 0$ will be a subject of a separate research.